\documentclass[aps, prl, reprint, showpacs, twocolumn, 10pt, groupedaddress,noeprint]{revtex4-2}

\usepackage{graphicx, amsmath, dsfont, dcolumn, bm, times, xcolor, braket, amssymb,natbib}

\begin{document}

\title{Five-terminal quantized transconductance originating from symmetric quantum fluctuations}

\author{K. Mertiri, Yuli V. Nazarov}
\affiliation{Kavli Institute of Nanoscience, Delft University of Technology, 2628 CJ Delft, The Netherlands}
\date{\today}

\begin{abstract}
We consider the renormalization of transport in a quantum contact by the external electromagnetic environment and show that weak quantum fluctuations from a symmetric environment can lead to a quantized transconductance in five-terminal contacts. This mechanism does not work for fewer terminals. We present an implementation of the environment and investigate several example quantum contacts where the quantization takes place. 
\end{abstract}

\maketitle

Quantum transport in nanostructures is conveniently described in terms of scattering of electron waves coming from and going to macroscopic terminals \cite{QTBook}. Interaction between electrons can, however, dramatically alter scattering. At low energy scales the most important contribution comes from interactions with the voltage fluctuations provided by the external electromagnetic environment, with the coupling strength characterized by the dimensionless impedance $z \equiv (G_Q/2) Z(\omega)$, $G_Q\equiv e^2/\pi \hbar$. The effect of interaction was first described in the context of tunnel junctions \cite{Tunnel1,Tunnel2,Tunnel3}. At $z \to \infty$, it leads to the Coulomb blockade, i.e. complete insulation of the contacts. Most interestingly, it was shown that the blockade persists at $z\ll1$ in the limit of small energy scales $E \sim \text{max}(k_BT, eV)$, as defined by the temperature, $T$, and voltage, $V$, across the junction. This behavior at low energies was later confirmed for 2-terminal quantum contacts of  arbitrary transparency \cite{Coulomb1,Coulomb2}.

In the scattering paradigm, the effect of electron-electron interactions in 2-terminal contacts at $z\ll1$ can be accounted for through the renormalization of the transmission coefficient $T_i$ of channel $i$. Starting from an upper energy cutoff, $E_\text{cut}$, quantum fluctuations act to iteratively rescale $T_i$. The resulting renormalization equation expresses the energy dependence of the effective transmission coefficients:
\begin{equation}\label{eq:tunnelRenormalization}
    \frac{dT_i}{d\xi} = -2zT_i(1-T_i),
\end{equation}
$\xi = \ln(E_\text{cut}/E)$. The only stable fixed point of this flow equation is the insulating point where all $T_i =0$. Thus the equation describes the gradual suppression of transmission as $\xi \to \infty $ ($E\to 0$). The equivalent equation for a general $z$ is not known for arbitrary transmissions.

Recently, we extended the study of environment-induced renormalization to multi-terminal quantum contacts \cite{quantizedTransconductance}. Remarkably, we found that quantum fluctuations coming from environments that break time-reversal symmetry can lead to quantized transconductance (QTC) at low energies. The quantization arises from an integer number of fully transmitting chiral channels. While the effect is phenomenologically similar to the Integer Quantum Hall Effect \cite{Buttiker}, it is of a different origin. QTC has also been predicted in junctions of three strongly interacting quantum wires described by the Tomonaga-Luttinger liquid theory \cite{LL}. The renormalization in TL models is qualitatively similar but quantitatively different than that in the environment model. There are also other proposals to realize quantized transconductance based on superconducting devices \cite{QT1,QT2,QT3}.

In this Letter, we show that electromagnetic environments where time reversibility is preserved, can also give rise to quantized transconductance. We show that the minimum number of terminals for this is five. We design a concrete environment suitable for this effect and demonstrate the QTC for several illustrative quantum contacts. The time-reversal symmetry breaking required for QTC comes from the contact properties. In our examples, the symmetry is broken by the magnetic flux penetrating the contact.

\begin{figure}
    \centering
    \includegraphics[width=0.8\linewidth]{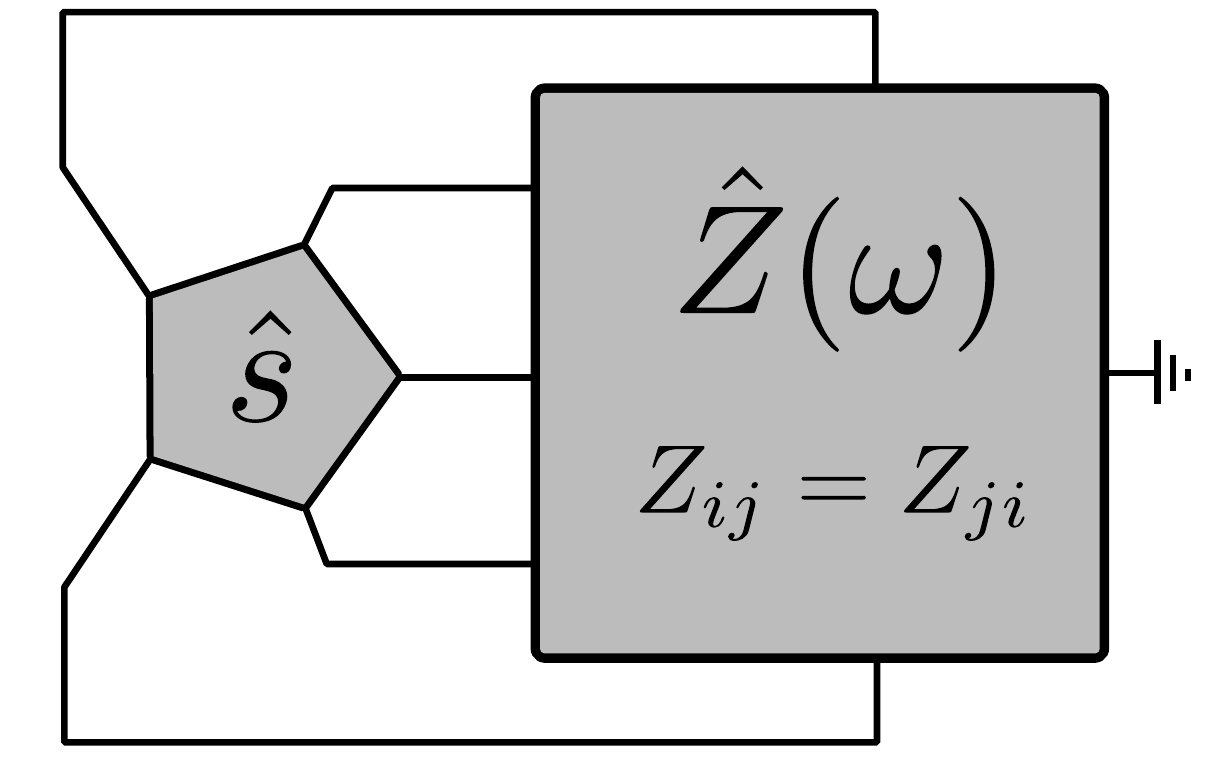}
    \caption{The setup under consideration. A five-terminal quantum contact, characterized by the scattering matrix $\hat s$, is connected to a time-reversible electromagnetic environment described by the frequency-dependent $5\times5$ symmetric impedance matrix $\hat Z(\omega)$. The high-frequency fluctuations in the environment can develop QTC in the contact. To uncouple the contact and environment at very low frequencies, which is required for transport measurements, the connection is realized via large capacitors not shown in the figure.}
    \label{fig:setup}
\end{figure}

The setup we consider is shown in Fig.~\ref{fig:setup}. A five-terminal quantum contact is connected to a time-reversible electromagnetic environment, which is generally described by a {\it symmetric} $5 \times 5$ impedance matrix $\hat Z (\omega)$. We assume that the environment is Ohmic, so that the  dimensionless impedance  matrix $\hat z \equiv (G_Q/2) \text{Re} \hat Z$ does not depend on frequency in a wide frequency range. In addition, we assume small $z\ll1$, and find the effective energy dependent scattering amplitudes between channels $i$ and $j$ to obey the following renormalization equation \cite{quantizedTransconductance}:
\begin{equation}\label{eq:renormalization}
    \frac{ds_{ij}}{d\xi} = z_{ji}s_{ij} - \sum_{k,l} z_{kl}s_{ik} (\hat s^\dag)_{kl} s_{lj},
\end{equation}
where, without loss of generality, we assume a separate terminal for each channel.

The renormalization equation admits the following fixed points: $\hat s_{ij}^{(0)} = e^{i\theta_i'} P_{ij} e^{i\theta_j}$, where  $\hat P$ is a permutation matrix (that is, one with exactly one $P_{ij} = 1$ non-zero element in each row and each column) and  $\theta_{i}'$, $\theta_{j}$ are arbitrary phases. The stability condition for a permutation fixed point reads:
\begin{equation}\label{eq:gstability}
    (\hat{z}\hat{P})_{ii} + (\hat{z}\hat{P})_{jj} - (\hat{z}\hat{P})_{ij} - (\hat{z}\hat{P})_{ji} >0 \quad \forall\, i,j. 
\end{equation}
In addition, the impedance matrix must be positively defined (i.e. the environment is stable and passive). The two conditions are always met for the trivial insulating point $\hat P = \hat 1$. In \cite{quantizedTransconductance} we found non-trivial stable permutation points for an environment with broken time-reversal symmetry ($\hat z$ is asymmetric), which correspond to QTC. In this Letter, we examine the realization of QTC points for a time-reversible environment with a symmetric $\hat z$.

Permutation matrices can be decomposed into disjoint cycles, so that condition \eqref{eq:gstability} is imposed on $i,j$ within the same or different cycles. We found that cycles of length 2 are always unstable, whereas cycles of length 3 and 4 are stable only if $\hat z$ is asymmetric. So that, only cycles of length 5 or more can be stable in symmetric environments. In what follows, we concentrate on cycles of length 5 in five-terminal contacts. There are many possible design choices for a $5 \times 5$ $\hat z$. For simplicity, we stick to a cyclically symmetric environment, such that $z_{i,i\pm n} \equiv Z_n$, $\forall  i$ ($n<3$), and the environment is completely characterized by three parameters $Z_{0,1,2}$. For such environment, we show that there are four potentially stable QTC fixed points. Two of them, denoted by $P_{\pm1}$, correspond to transmission between nearest-neighbor terminals, either in clock- or counterclockwise direction, so the channels draw a pentagon. The other two, $P_{\pm2}$, correspond to transmission between next-nearest-neighbor terminals, so the channels draw a five-pointed star. 

\begin{figure}
    \centering
    \includegraphics[width=0.75\linewidth]{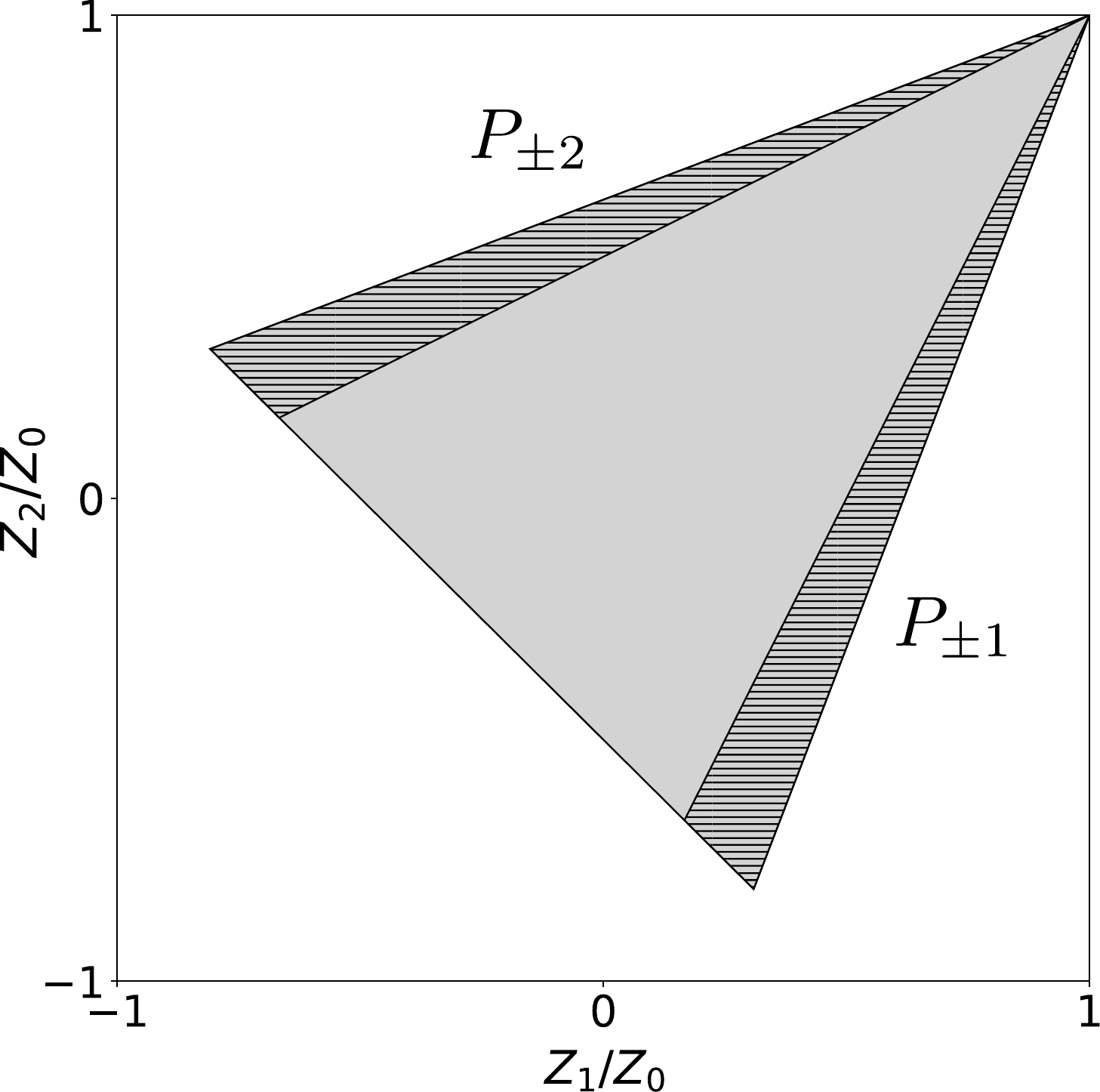}
    \caption{The region of positively defined impedance matrices (gray) overlapped by the region of stability (hatched)  for $P_{\pm1}$ and $P_{\pm2}$ in the space of matrix elements $Z_i$. The stability region corresponds to amplifying Ohmic environments.}
    \label{fig:stability}
\end{figure}

Let us map the stability conditions (Fig.~\ref{fig:stability}) for $Z_{1,2}/Z_0$. The trivial reasoning on the positivity of $\hat z$ involving one terminal gives $Z_0 >0$, and involving two terminals gives $|Z_{1,2}| < Z_0$. This yields a square in  Fig.~\ref{fig:stability}. More accurate consideration involving more terminals gives 3 conditions on the 3 eigenvalues of $\hat z$, $Z_0 + 2 Z_1 + 2 Z_2>0$, $Z_0 \pm (\sqrt{5}\mp1)/2 \ Z_1 \mp  (\sqrt{5}\pm1)/2 \ Z_2 >0$, that carve a gray triangle within the square. The non-trivial stability conditions \eqref{eq:gstability} for the QTC fixed points described read:
\begin{equation}\label{eq:stability}
\begin{aligned}
    P_{\pm1}:&\quad 2Z_1 > Z_0 + Z_2,\\
    P_{\pm2}:&\quad 2Z_2 > Z_0 + Z_1.
\end{aligned}
\end{equation}
These two lines define the mutually exclusive hatched regions of interest on two sides of the triangle.  

Let us discuss the realization of such environment. In general, the environment must be (a) symmetric, (b) amplifying, (c) passive, (d) broadband and (e) must allow for low-frequency measurements on the contact. Criteria (c) and (d) are required by the renormalization technique. Criterion (e) can be met by separating the environment and contact with large capacitor filters. The important criterion is (b). Most circuits assembled from passive elements are non-amplifying at low frequency: if a current is run between two contacts, the voltages of all other contacts lie between the voltages of the two. This is in contradiction with the general stability conditions \eqref{eq:gstability} for non-trivial points: If we set $j \equiv P^{-1}(i)$ in \eqref{eq:gstability}, the condition becomes: $z_{iP(i)} +z_{P^{-1}(i)i} - z_{ii} -z_{P^{-1}(i)P(i)}$. This is equivalent to the requirement that upon passing a current through terminals $i \to P(i)$, the voltage $V_{P^{-1}(i)}>V_i$.

\begin{figure}
    \centering
    \includegraphics[width=\linewidth]{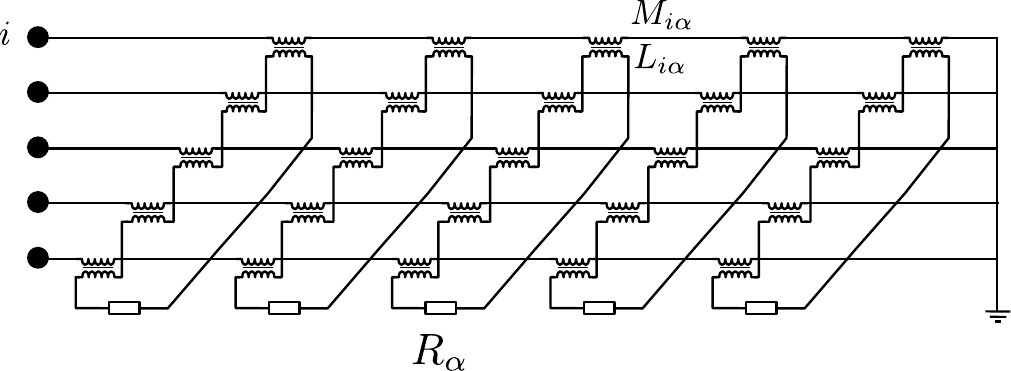}
    \caption{Implementation of the electromagnetic environment. General design. In each of the 5 loops labeled by $\alpha$, the self inductances of the ideal transformers $M_{i\alpha},L_{i\alpha}$ are chosen to correspond to $\psi^{(\alpha)}_i$, an eigenvector of $\hat{z}$. The resistor $R_\alpha$ gives the corresponding eigenvalue. A cyclically symmetric Ohmic environment is obtained when $\psi^{(\alpha)}_i$ correspond to Fourier modes, and $R_1=R_4$, $R_2=R_3$.}
    \label{fig:circuit}
\end{figure}

These criteria can all be satisfied with a proper design, (b) requiring (on-chip) transformer networks \cite{networkbook}. We propose an implementation of the most general five-terminal cyclically symmetric environment. The design is shown in Fig.~\ref{fig:circuit}. Ideal transformers with self-inductance $M_{i\alpha},L_{i\alpha}$ couple terminal $i$ to resistor $\alpha$ with resistance $R_\alpha$. At large frequencies $\omega \gg R/L$, with $R, L$ the scale of resistance and inductance respectively, the real part of the impedance matrix is Ohmic and can be amplifying. There is a large range of possible choices for the parameters of this design. Here, we set $M_{i\alpha} = \sum_j L_{j\alpha}$, $\forall i$, so that inductances $L_{i\alpha}$ are directly related to the eigenvectors $\Psi_i^{(\alpha)}$ of the impedance matrix, and $R_\alpha$ equal to its eigenvalues.  More specifically:
\begin{equation}
	\Psi^{(\alpha)}_i  = p_{i\alpha} \sqrt{\frac{L_{i\alpha}}{\sum_j L_{j\alpha}}},
\end{equation}
with $p_{i\alpha} = \pm 1$ the polarity of the respective transformer. The eigenvectors of all cyclically symmetric matrices are the same and equal the Fourier modes, $\Psi^{(\alpha)}_i \propto \cos( 2\pi \alpha i/5)$ for $\alpha=0,1,2$, and $\Psi^{(\alpha)}_i \propto \sin(2\pi \alpha i/5)$ for $\alpha=3,4$. The cyclically symmetric environment is realized if $R_1=R_4$, $R_2=R_3$, $R_\alpha$ corresponding to the eigenvalues of $\hat{z}$. Simpler designs could also provide stable QTC points with fewer transformers, however, we prefer a more flexible one.

With the circuit implemented, we can achieve the stable QTC fixed points. However, the insulating fixed point is also stable. Upon renormalization, an `initial' high-energy $5\times 5$ scattering matrix $\hat s_\text{in}$ flows in the limit of low energy to either one of the two stable QTC points or the insulating point. To find out the probability for a scattering matrix to flow to the QTC points, we draw them at random from the  circular unitary ensemble, and solve the flow equation numerically. The cyclically symmetric impedance matrix we made use of in this and all following numerical calculations was chosen to sit well within the stability region of the corresponding QTC points.  For $P_{\pm 1}$, we work with the environment ${\cal E}_1$ where $R_0,R_1,R_2 =z(2/G_Q)(4.19, 1.80, 0.04)$ and for $P_{\pm 2}$, we work with  ${\cal E}_1$ where $R_0,R_1,R_2 =z(2/G_Q)(4.19, 0.04, 1.80)$, that is, $Z_{1,2}$ are exchanged.

We found the probability to be of the order of $10^{-5}$. While a rather discouraging result, it is easy to understand why the probability is so small. Five terminals is crucial for the effect, and the matrices where most scattering is dominated by fewer terminals should flow to the insulating point. We show that simple engineering of the initial scattering matrix suffices to achieve the QTC fixed points. 

First of all, the scattered waves should ideally be uniformly distributed among all five terminals. This implies that the initial scattering matrix must not have large reflection amplitudes for the conducting channels. Secondly, time reversibility must be broken: since this is preserved by the flow equation, time-reversible $\hat{s}_\text{in}$ all flow to the insulating point. The first requirement is fulfilled by a design that makes possible propagation from any terminal to any other. The second requirement implies scattering trajectories sensitive to magnetic flux, that is, loops. There must also be sufficient probability to travel across such trajectories. 

The required contacts can be implemented in a 2D electron gas (2DEG) hosted in semiconductor heterostructures, an approach pioneered in Ref.~\cite{vanWeesquantized}. By using top electrodes to deplete the 2DEG, one can define a scattering region, while additional gates provide control over the scattering potential, and thus the scattering matrix, and dynamical phases \cite{Buchholz}. This approach enables the realization of complex devices based on 2DEG-defined beam splitters \cite{beamsplitter}, as well as loops \cite{leturcq2006asymmetries,fuhrer2004quantum,Buchholz} and multi-terminal geometries \cite{pankratova2020multiterminal,bachsoliani2017mesoscopic}. Beyond 2DEG-based devices, similar structures can be realized using suspended graphene \cite{ki2013high}, or through more exotic approaches involving nanowires \cite{kahn2025phase}.

\begin{figure}
    \centering
    \includegraphics[width=\linewidth]{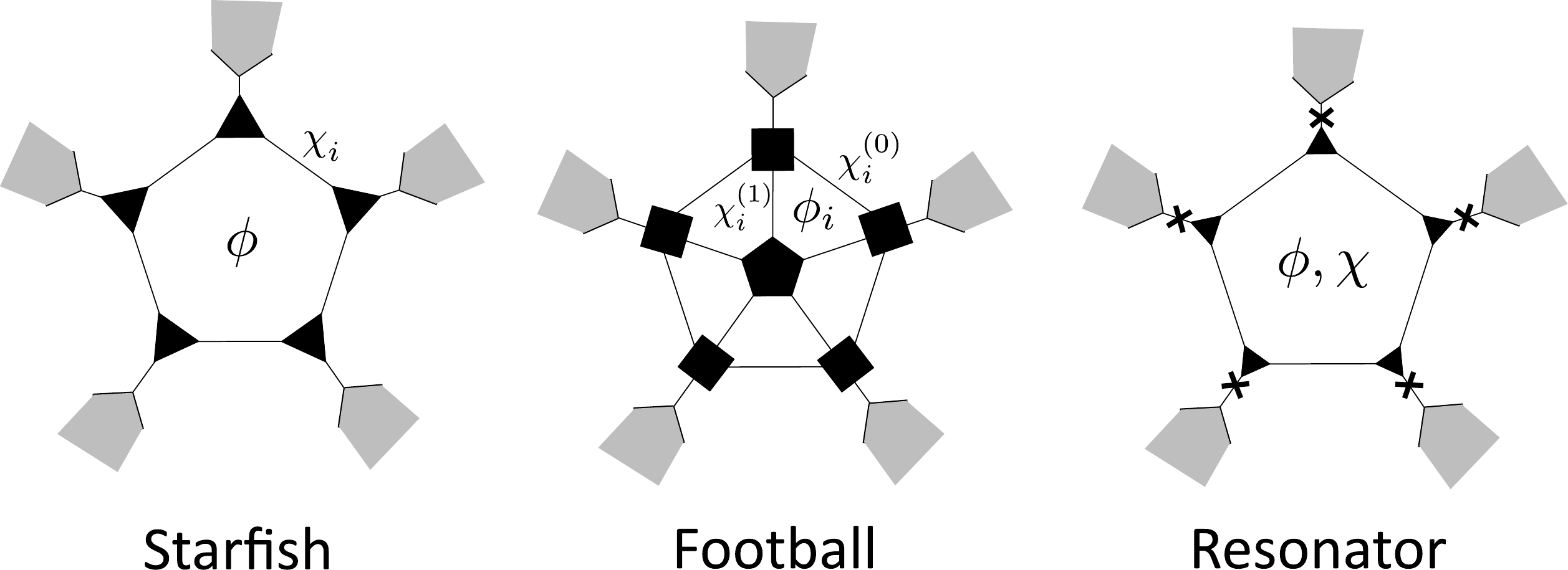}
    \caption{Three setups proposed to demonstrate the QTC fixed points are built from beam splitters forming loops. Gray polygons represent the terminals, lines represent the propagation channels, and black $n-th$ order polygons represent the beam splitters, corresponding to a unitary $n \times n$ matrix.  In the Starfish setup, electrons follow the loop unless exiting to the terminals underway. In the Football setup, the central beam splitter provides more chances to explore the loops. The beam splitters in the Resonator have a low entrance transparency, this is indicated by crosses. If a resonance condition is achieved, this guarantees electron trapping in the loop, so that it can explore all terminals.}
    \label{fig:setups}
\end{figure}

Based on these principles, we design three different setups building $5\times5$ scatterers from beam splitters realizing the loops. We dub them Starfish, Football, and Resonator (Fig.~\ref{fig:setups}). Below we discuss the setups and the parameters required to achieve the QTC points.

\begin{figure}
    \centering
    \includegraphics[width=\linewidth]{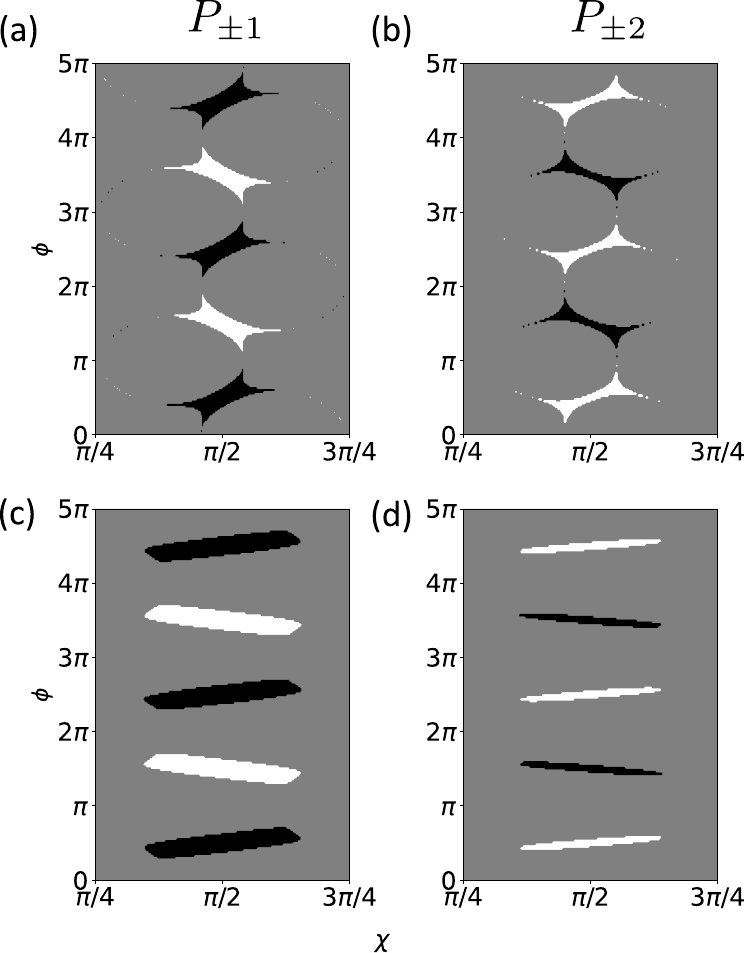}
    \caption{The Starfish setup. Convergence regions to QTC points of opposite chirality $\pm$ are shown in black and white, respectively, the gray background indicating the flow to the insulating point. For plots (a) and (c), the environment is ${\cal E}_1$ favoring $P_{\pm1}$, while in (b) and (d) the environment is ${\cal E}_2$ favoring $P_{\pm2}$. For plots (a) and (b), all dynamical phases are chosen to be equal $\chi_i \equiv \chi$, whereas for plots (c) and (d) the choice is $\chi_0 \equiv \chi$, with the rest $\chi_{i\ne 0} = \pi/2$.}
    \label{fig:starfish}
\end{figure}

The Starfish setup is composed of five 3x3 fully-symmetric beam splitters, with scattering matrix:
\begin{equation}
\hat s_3 = \frac{2}{3}
	\begin{pmatrix}
        	-1/2 & 1 & 1\\
        	1 & -1/2 & 1\\
        	1 & 1 & -1/2
	\end{pmatrix}.
\end{equation}
The splitters are connected to form a loop that encloses the flux $\Phi = (\hbar/e)  \phi$. Electrons propagating from the splitter $i$ to $i+1$ (from $i+1$ to $i$) acquire magnetic phase $\phi/5$(-$\phi/5$), and dynamical phase $\chi_i$. The resulting scattering matrix $\hat s_\text{in}(\phi, \{\chi_i\})$ arises from interference between many scattering events as the electron travels through the nanostructure, and, crucially, is a function of the phases. We numerically solve the renormalization equation starting with $\hat s_\text{in}$ to the limit of low energies and determine the fixed point it converges to. This allows us to draw the convergence regions for the QTC fixed points in the space of $\phi$ and $\chi$. The results are shown in Fig.~\ref{fig:starfish}. Plots (a) and (c) on the left show the convergence regions to $P_{\pm1}$ for the environment ${\cal E}_1$, while plots (b) and (d) on the right illustrate the convergence to $P_{\pm2}$, for ${\cal E}_2$. In plots (a) and (b), we set all dynamical phases to be the same $\chi_i \equiv \chi$. At zero flux, $\chi = \pi/2$ minimizes the reflection coefficient. The convergence regions are therefore centered around  $\chi = \pi/2$. As flux increases, the asymmetry of $\hat s_\text{in}$ favors either $P_1$ or $P_{-1}$. In plot (b), the convergence regions are narrower, since in the Starfish setup the transmission to the nearest-neighbor terminals generally is larger than that to the next-nearest-neighbor terminals. Plots (c) and (d) show the convergence regions as a function of the dynamical phase between splitters 0 and 1, $\chi_0 \equiv \chi$, and flux. All other dynamical phases are set to the optimal value of $\pi/2$. This is why the regions in (c) are bigger than in (a). We observe peculiar shapes of the convergence regions for the symmetric configurations in (a) and (b). The shapes make cusps from where thin quickly tapering filaments extend over some curves in the plane. Since the filaments are very thin, we had difficulty to resolve them with proper precision. These features are not seen for the less symmetric configurations in (c) and (d), so we regard those as artifacts of the high symmetry and do not pay special attention to them.  

\begin{figure}
    \centering
    \includegraphics[width=\linewidth]{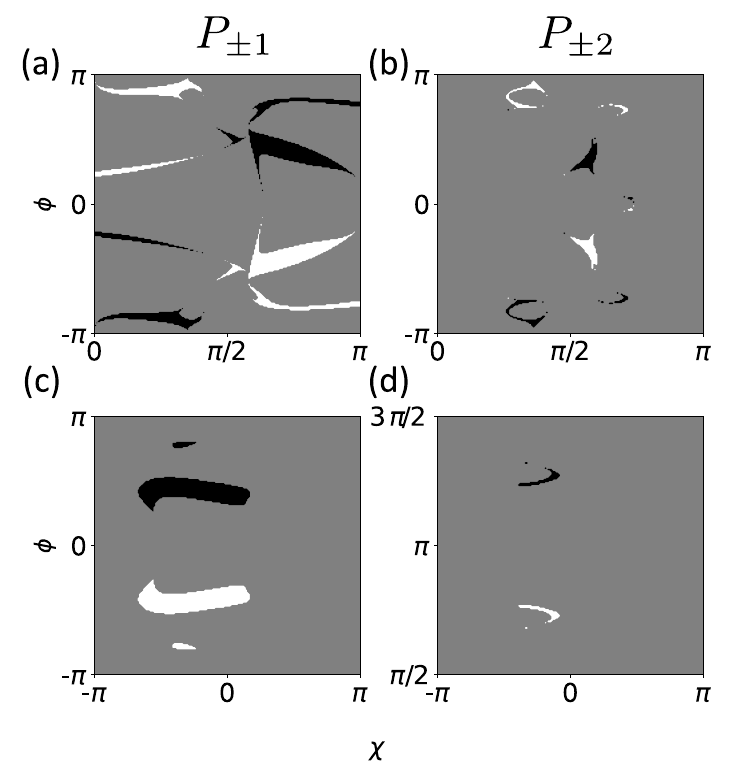}
    \caption{The Football setup. Convergence diagram. Plots (a) and (c) show the convergence regions to $P_{\pm1}$ (black and white) for the environment ${\cal E}_1$ , while plots (b) and (d) show this for $P_{\pm2}$ and environment ${\cal E}_2$. In plots (a) and (b) we tune the flux and all dynamical phases $\chi^{(0)}_i \equiv \chi$, with $\chi^{(1)}_i =0$, while in plots (c) and (d) we tune the flux and the dynamical phase of one channel, $\chi^{(0)}_0 \equiv \chi$, with $\chi^{(0)}_{i\ne0} = -\pi/3$, $\chi^{(1)}_i = -\pi/2$.}
    \label{fig:football}
\end{figure}

An obvious drawback of the Starfish setup is the chance for an electron to escape the terminals while traversing the loop, so that the probability to complete the loop is relatively small. We partially circumvent this in the Football setup, composed of five 4x4 reflectionless beam splitters, with scattering matrix
\begin{equation}
\hat s_4 = \frac{1}{\sqrt{2}}
	\begin{pmatrix}
        	0 & 1 & 0 & 1\\
        	1 & 0 & -1 & 0\\
        	0 & -1 & 0 & 1\\
        	1 & 0 & 1 & 0
	\end{pmatrix},
\end{equation}
connected to each other in a loop and to a central 5x5 reflectionless beam splitter, with a cyclically symmetric scattering matrix $\hat s_5$ with elements $(\hat s_5)_{i,i} = 0$, $(\hat s_5)_{i,i\pm1} =  \exp(-i\pi/3)/2$, and $(\hat s_5)_{i,i\pm2} = \exp(i\pi/3)/2$, $\forall i$. This way, electrons explore more loops before escaping to the terminals. For all illustrations, we set the same flux $\Phi = (\hbar/e)  \phi$ in each loop. The channels connecting the 4x4 splitters acquire dynamical phases $\chi^{(0)}_i$, while channels connecting to the 5x5 splitter acquire dynamical phases $\chi^{(1)}_i$. We take two sets of parameters to illustrate. The convergence diagram is shown in Fig.~\ref{fig:football}. In plots (a) and (b) the flux and dynamical phases $\chi^{(0)}_i \equiv \chi$ are tuned, with $\chi^{(1)}_i=0$.  In plots (c) and (d) we tune the flux and the dynamical phase along one of the channels $\chi^{(0)}_0\equiv \chi$, with the rest set to $\chi^{(0)}_{i\ne0} = -\pi/3$; $\chi^{(1)}_i = -\pi/2$. Again, we compute the convergence regions for two different environments ${\cal E}_{1,2}$. In comparison with Fig. \ref{fig:starfish} the phase diagram is more complex, with the QTC convergence regions taking a bigger part of the phase space. 

\begin{figure}
    \centering
    \includegraphics[width=\linewidth]{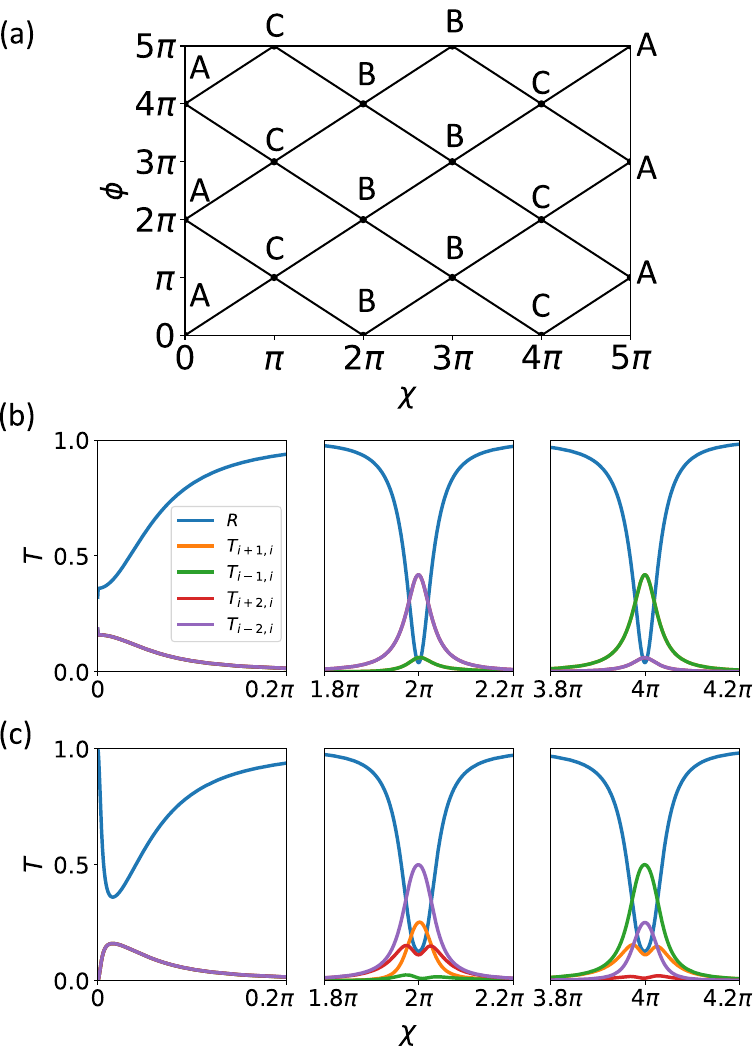}
    \caption{The Resonator setup. (a) The positions of the resonance levels vs dynamical, $\chi$, and magnetic, $\phi$, phases. There are three types of level crossings, A, B and C, as labeled. (b) Initial transmission coefficients in the vicinity of the resonance crossings for $\phi=0$. (c) Initial transmission coefficients for $\phi=0.01$ of the order of the small entrance transmission coefficient. Crossings A, B and C, are characterized by equal, next-nearest-neighbor and nearest-neighbor transmission, respectively.}
    \label{fig:crossings}
\end{figure}

In both the Starfish and Football setups electrons leave the contact relatively fast, which limits flux sensitivity and the number of terminals explored. This can be circumvented with the Resonator setup, where the beam splitters have small entrance/exit amplitudes and are almost reflectionless within the channels that connect them to each other. Once an electron enters the loop, it spends a long time there, allowing it to feel the flux and explore all terminals. This only occurs at a resonant condition: the Fermi energy coinciding with one of the resonant energy levels formed in the loop. The resonance condition is fulfilled at $\phi \pm \chi = 2\pi N$, with $N$ an integer and $\phi$, $\chi$ the magnetic and dynamical phase acquired upon traversing the loop once. The reflection coefficient is smallest at the crossings between two resonances. The resonance positions as a function of phases are shown in Fig.~\ref{fig:crossings}(a). There are three types of crossings labeled A, B and C. The difference between them is best illustrated in Fig.~\ref{fig:crossings}(b), where we plot the transmission coefficients $T_{i,j}$ as a function of $\chi$ at $\phi = 0$. At an A crossing, the transmission coefficients are all the same, while at a B crossing , $T_{i\pm1,i}<T_{i\pm2,i}$, and  at a C crossing,  $T_{i\pm1,i}>T_{i\pm2,i}$. Figure~\ref{fig:crossings}(c) illustrates the transmission coefficients near the crossings at finite small flux $\phi = 0.01$ (of the order of the small entrance transmission coefficient).  Near an A crossing, the flux simply shifts the resonance introducing no asymmetry. The flow to the insulating fixed point is therefore expected near these crossings. The flux does introduce asymmetry near B and C crossings, enabling the flow to $P_{\pm 2}$ and $P_{\pm 1}$, respectively, as confirmed by numerical calculations. 

We explore the convergence to the QTC points in the vicinity of a C crossing: $\chi,\phi = \pi$ and a B crossing: $\chi = 2\pi,\phi = 0$ choosing the environments ${\cal E}_1$, ${\cal E}_2$, respectively. In computations, tunneling to and from the leads is modeled by the 3x3 scattering matrix:
\begin{equation}
    \hat s_r = 
    \begin{pmatrix}
        \cos(2\delta) & \frac{i}{\sqrt 2} \sin(2\delta) &  \frac{i}{\sqrt 2} \sin(2\delta)\\
       \frac{i}{\sqrt 2} \sin(2\delta) & -\sin^2 \delta & \cos^2 \delta\\
         \frac{i}{\sqrt 2} \sin(2\delta) & \cos^2 \delta &  -\sin^2 \delta
    \end{pmatrix},
\end{equation}
and the small entrance transmission coefficient $T_{\rm s} = 2\delta^2$. 

\begin{figure}
    \centering
    \includegraphics[width=\linewidth]{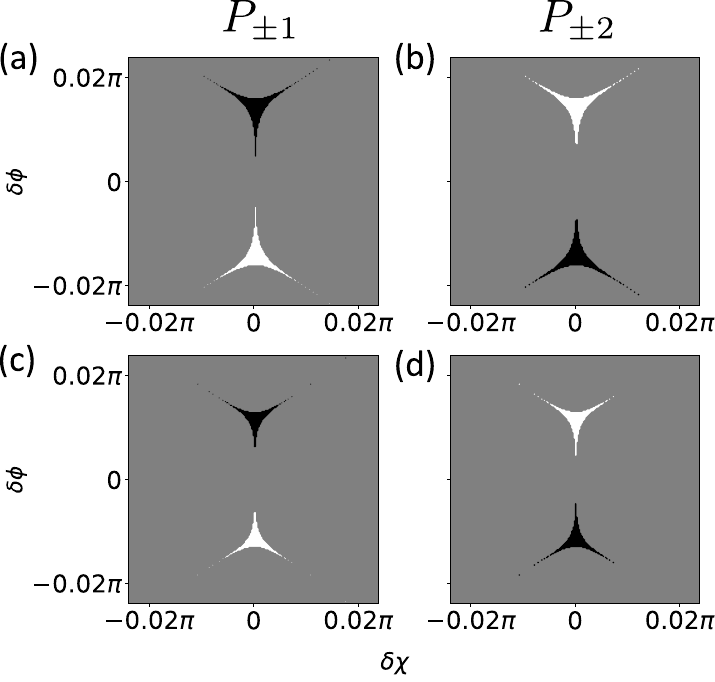}
    \caption{The Resonator setup. Convergence diagram. Plots (a) and (c) show the convergence to $P_{\pm1}$ in the vicinity of a C crossing, while plots (b) and (d) show the convergence to $P_{\pm2}$ near a B crossing. The flux and dynamical phase are counted from the crossing positions. The scale of the plots is the small transmission coefficient $T_{\rm s}$. Concretely, $T_{\rm s} = 0.04$ for (a) and (b) and $T_{\rm s} = 0.03$ for (c) and (d).}
    \label{fig:resonator}
\end{figure}

The convergence regions are shown in Fig.~\ref{fig:resonator}. Plots (a) and (c) correspond to the vicinity of a C crossing (where $P_{\pm1}$ are stable), while (b) and (d) correspond to the vicinity of a B crossing ( where $P_{\pm2}$ are stable), $\delta \phi, \delta \chi$ being the deviations of the magnetic and dynamical phase from the crossing position. In all plots, the scale of the convergence regions in both $\delta \phi$ and $\delta \chi$ is determined by $T_{\rm s}$. The centers of the black and white regions are separated by $\Delta \phi \approx 2T_s$, and their width is correspondingly $\Delta \phi, \Delta \chi \approx T_{\rm s}/4$. We also notice the shape peculiarities of the region: similar to the shapes in Figs.~\ref{fig:starfish}(a) and (b), there are filaments extending and tapering along the lines $\delta \chi = 0$, and two slanted lines (not parallel to the resonant level positions). We regard it as an artifact of the high symmetry and do not investigate these in detail.
The plots on the left (for $P_{\pm 1}$) and on the right (for $P_{\pm 2}$) have the same shape in the limit of small $T_{\rm s}$, the relative difference being of the order of $T_{\rm s}$. This is because of almost reflectionless propagation along the loop. To make the scaling with $T_{\rm s}$ visible, we use slightly different $T_{\rm s}$ in (a), (b) ($T_{\rm s} =0.04$) and in (c), (d) ($T_{\rm s} =0.03$). 

\begin{figure}
    \centering
    \includegraphics[width=\linewidth]{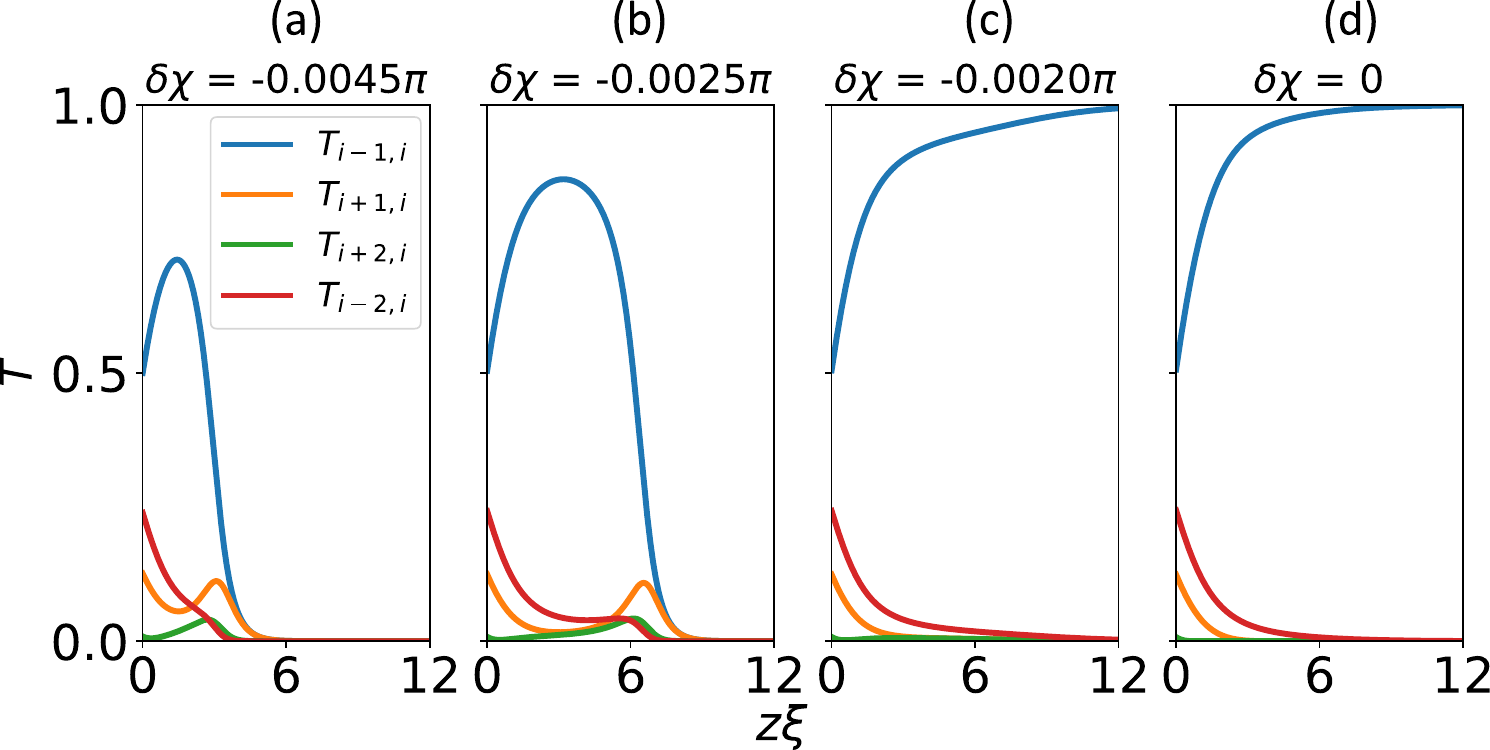}
    \caption{The Resonator setup. The renormalization flow for a selection of parameters in Fig.~\ref{fig:resonator}(a). We plot the transmission coefficients versus the renormalization parameter $z\xi$.}
    \label{fig:evolution}
\end{figure}

In Fig.~\ref{fig:evolution}  we present the renormalization flow near a C crossing with the parameters corresponding to Fig.~\ref{fig:resonator}(a). The flux is chosen in the middle of the white region $\delta \phi = -0.015\pi$, while dynamical phase settings are (a) deep in the gray region, (b) very close to the boundary from the gray side, (c) very close to the boundary from the white side, and (d) in the middle of the white region.  For all settings, we plot the transmission coefficients as a function of $z\xi$. $T_{i\pm n,i}$ are the same for all $i$, and that does not change with the flow. In the leftmost plot (a), all coefficients go to 0 in the low energy limit, approaching the insulating point. As the dynamical phase approaches the stability threshold, the flow becomes slower. Above the threshold, $T_{i-1,i}$ quickly flows to 1, approaching $P_{-1}$. The traces in (b) and (c) are very similar in an interval $z\xi = [0,4]$. 

The Resonator setup looks promising, yet two points need to be clarified. Firstly, the small entrance transmission coefficient results in a Coulomb blockade, which we formally did not account for. We think it is not important, since the Coulomb blockade only shifts the resonant level positions and thus can be compensated with proper gates. Secondly, the resonance implies the energy dependence of the scattering matrix on an energy scale of the order of the resonant level width. By virtue of our renormalization approach, this energy scale should be {\it larger} or of the order of the upper cutoff energy for the renormalization. This condition can always be met by choosing very short contacts.

Finally, an important note about the setups presented: they do not have to be realized literally, by carving actual loops and channels. The only important condition is the exploration of the whole nanostructure by the scattered electrons. For instance, the loops can be replaced by continuous propagation regions, like a disk connected to five terminals. All flux effects except the periodicity should be preserved in these setups.

\textit{Conclusion and Discussion}. In conclusion, we have demonstrated that quantum fluctuations of a time-reversible Ohmic environment may lead to quantized transconductance in the limit of low energies. This is the result of the stability of non-trivial fixed points in the renormalization flow. Rather ironically, one needs at least five-terminal quantum contacts and environments to observe this phenomenon. We concentrated on a cyclically symmetric environment and derived the stability criteria for the QTC fixed points in terms of the characteristic impedances. We have shown how to realize the required environment from resistors and transformers. Importantly, we found examples of quantum contacts that converge to the QTC fixed points upon renormalization. This facilitates the experimental observation of the phenomenon.

We note that this topic motivates further research. Our consideration was limited to the simplest case of one channel per terminal. It is interesting to investigate the case of multiple channels, in particular the semiclassical case of many channels, to extend to bigger nanostructures. The permutation group suggests the existence of even more complex QTC fixed points, and it would be interesting to identify those in such nanostructures.  Furthermore, we restricted our analysis to small dimensionless impedances of the environment. Assessing the situation at bigger impedances is a challenging theoretical problem. In principle, the QTC fixed points must remain fixed at any impedance, while the stability criteria can be very different.

\textit{Acknowledgments}. We are indebted to E. S. Samuelsen, F. von Oppen, C. Sch\"{o}nenberger,  and N. Andrei for interesting discussions of several aspects of this topic. This research is supported by Delft University of Technology. 

\textit{Data availability}. The data that support the findings of this article are openly available at \cite{data}.

\bibliography{paper-bib}
\end{document}